\documentclass[ superscriptaddress]{revtex4-2}
\usepackage{graphics,epsfig}
\usepackage{epstopdf}
\usepackage{graphicx}
\usepackage{dcolumn}
\usepackage{amsmath}
\usepackage{epstopdf}
\usepackage{caption}
\usepackage{subcaption}
\usepackage{hyperref}
\begin{document}
%opening
\title{Extended thermodynamics and $P-v$ Criticality of Kalb-Ramond  black  hole coupled with nonlinear electrodynamics}

\author{Dharm Veer Singh}
\email{veerdsingh@gmail.com}
\affiliation{Department of Physics,
Institute of Applied Science and Humanities, GLA University, Mathura - 281406, Uttar Pradesh, 
India \footnote{Visiting Associate: Inter-University Center of Astronomy and Astrophysics (IUCAA), Pune}}
\affiliation{School of Physics, Damghan University, P.O. Box 3671641167, Damghan, Iran}

 \author{Sudhaker Upadhyay\footnote{Corresponding author}\footnote{Visiting Associate at IUCAA Pune, India}}
\email{sudhakerupadhyay@gmail.com}
\affiliation{Department of Physics, K.L.S. College, Magadh University, Nawada 805110, India}
 \affiliation{School of Physics, Damghan University, P.O. Box 3671641167, Damghan, Iran}
\author{Prosenjit Paul}
\email{prosenjitpaul629@gmail.com}
\affiliation{Indian Institute of Engineering Science and Technology (IIEST), Shibpur-711103, WB, India}
\author{Kairat Myrzakulov}
\email{krmyrzakulov@gmail.com}
\affiliation{Department of General \& Theoretical Physics, L. N. 
Gumilyov Eurasian National University, Astana, 010008, Kazakhstan}
\begin{abstract}
We present an exact black hole solution in anti–de Sitter (AdS) spacetime with a Kalb-Ramond field coupled to nonlinear electrodynamics (NLED), characterized by mass, magnetic monopole charge, and Lorentz-violating parameters. The geometry admits two horizons (inner and outer) that coalesce into a degenerate horizon at a critical monopole charge. Beyond this critical point, no black hole solutions exist. In the limit of vanishing Lorentz-violating parameters, the solution reduces to the modified Kalb–Ramond and Bardeen black holes, while suitable parameter choices reproduce the Reissner–Nordstr\"om–AdS and Schwarzschild-AdS geometries. We analyze the thermodynamics of the solution by computing the Hawking temperature, entropy, specific heat, and Gibbs free energy. The NLED source introduces nontrivial modifications: the Hawking temperature displays non-monotonic behavior with possible local extrema, the entropy deviates from the standard area law, and the specific heat may assume negative values, signaling thermodynamic instabilities. The Gibbs free energy exhibits swallow-tail structures, indicative of first-order phase transitions. Furthermore, we derive the first law of black hole thermodynamics in the extended phase space, together with the Smarr relation, and confirm their validity for the Kalb–Ramond black holes with NLED sources. Our findings highlight the rich thermodynamic structure induced by Lorentz-violating effects and nonlinear electrodynamics in AdS black hole backgrounds.
\end{abstract}

\maketitle

\section{\label{sec:level1}Introduction}

Lorentz symmetry forms a foundational aspect of the two leading frameworks in physics, general relativity and the standard model. Although both have remarkably explained numerous experimental results, many researchers suspect that Lorentz symmetry might not hold at extremely high energy scales. This potential breakdown appears in several advanced theoretical approaches, including string theory, non-commutative field theory, and others \cite{01,02,03,04,05,06}. The symmetry may undergo spontaneous or explicit violation. 

A violation of local Lorentz symmetry at the particle level may also arise due to a rank-two antisymmetric tensor field, commonly referred to as the Kalb-Ramond field \cite{07}.
In string-inspired theories, the Kalb-Ramond  field, arising from closed string excitations, is described by an antisymmetric rank-2 tensor. The corresponding rank-3 field strength associated with this Kalb-Ramond  field is often interpreted as spacetime torsion in the effective low-energy limit of higher-dimensional type IIB string theory \cite{7,8}. This torsional structure has been proposed to have deep connections with intrinsic particle spin, establishing a geometric bridge between string excitations and spin-related phenomena in field theory \cite{9,10}.

To address the issues of infinite self-energy and singularities arising in classical electrodynamics, particularly for point charges, a new approach was introduced through nonlinear electrodynamics (NLED). This framework was initiated by Born and Infeld in 1934, laying the foundation for regularised field theories \cite{Born:1934ji,Born:1934gh,Born:1935ap, Infeld:1936wzo, Infeld:1936frv}. NLED has since enabled the construction of black hole models that avoid central singularities. Motivated by astrophysical phenomena, various studies \cite{Ayon-Beato:2000mjt, Ayon-Beato:1999qin, Ayon-Beato:1998hmi} have explored such regular black holes within General Relativity and beyond, using NLED as the underlying matter field. Notably, Bardeen presented the first such solution, inspired by the earlier conceptual work of Gliner and Sakharov \cite{bardeen1968proceedings}. Based on the Bardeen proposal, there are many regular black hole solutions in refs \cite{Singh:2025svv, Kumar:2024zqf, Kumar:2024cnh, 
Sudhanshu:2024wqb, Singh:2024jgo, Kumar:2023cmo, Singh:2023czv, Hayward:2005gi, Singh:2022xgi, Kumar:2020bqf, Singh:2019wpu, Kumar:2018vsm, Ghosh:2018bxg, 2a,6a,Vishvakarma:2023tnl, sud00,sud01,ma,dvs20,1}. However, the presence of a cloud of strings turns the black hole solution singular for the Einstein gravity in the presence of non-linear electrodynamics \cite{sud,sud1,sud2}.

Bekenstein first introduced the notion that black holes possess entropy in 1972 \cite{Bekenstein:1973ur, Bardeen:1973gs}, who argued that this entropy is directly related to the surface area of the event horizon. Building on this idea, Hawking utilised quantum theory in 1974 to demonstrate that black holes can radiate energy, now known as Hawking radiation. Subsequently, Hawking and Page discovered a phase transition between $AdS$ black holes and pure thermal $AdS$ space \cite{Hawking:1982dh}, a result that carries significant consequences for the $AdS$/CFT correspondence and the study of quantum gravity.

In this paper, we derive exact black hole solutions in AdS spacetime incorporating a Kalb-Ramond  field coupled to  NLED, characterised by mass, magnetic monopole charge, and Lorentz-violating parameters $\gamma$ and $\lambda$. Our analysis identifies two distinct horizons: an inner and an outer horizon. As the monopole charge decreases, the outer horizon radius increases. The horizons merge at a critical charge value into a single degenerate horizon; no black hole solutions exist beyond this point. Without Lorentz-violating parameters, the solution reduces to the modified Kalb-Ramond  and Hayward solutions.
Additionally, specific parameter choices yield the RN–$AdS$ and Schwarzschild–$AdS$ black holes. We compute the Hawking temperature and examine the entropy, specific heat, and Gibbs free energy across various parameter configurations. Our findings indicate that the NLED source significantly alters the thermodynamic properties of the Kalb-Ramond  black hole, introducing non-monotonic behaviours in the Hawking temperature and deviations from the standard area law in entropy. The specific heat can become negative, signalling thermodynamic instability in certain regimes. We derive the first law of black hole thermodynamics in the extended phase space and confirm the validity of the Smarr relation, even in the presence of the NLED source. Our analysis also reveals that the Gibbs free energy exhibits non-monotonic behaviour, with local minima and maxima, indicating first-order phase transitions.

This paper is structured as follows. Section \ref{sec2} presents the derivation of black hole solutions in AdS spacetime, incorporating a Kalb-Ramond  field coupled to NLED. We examine the horizon structure and thermodynamic properties of these solutions. Section \ref{sec3} delves into the thermodynamics and stability of black holes, including the computation of Hawking temperature and entropy. We verify the first law of black hole thermodynamics and the Smarr relation within the extended phase space framework. To assess stability, we analyse the specific heat and Gibbs free energy. Section \ref{sec4} investigates the van der Waals-like phase transitions exhibited by these black holes, accompanied by the corresponding $G_{+}-T_{+}$ diagrams. Finally, section \ref{sec5} concludes the study by summarising the key findings and implications.

\section{Exact solution of Kalb-Ramond black hole}\label{sec2}
In this section, we investigate black hole solutions in the presence of a Kalb-Ramond  field coupled to  NLED within an  $AdS$ spacetime. We begin by considering the action for a self-interacting Kalb-Ramond  field that is non-minimally coupled to gravity, expressed in the form:
\begin{eqnarray}
{\cal {S}}&=&\int d^4x \sqrt{-g}\Big[\frac{R}{2\kappa}-\frac{1}{12}H_{\lambda\mu\nu}H^{\lambda\mu\nu}-V(B_{\mu\nu}B^{\mu\nu}\pm b_{\mu\nu}b^{\mu\nu})\nonumber\\
&+& \frac{1}{2\kappa}(\xi_2B^{\lambda\nu}B^{\nu}_{\mu}+R_{\lambda\mu}+\xi_3B^{\mu\nu}B_{\mu\nu}R)+L(F) - \frac{\Lambda}{\kappa}\Big],
\label{action}
\end{eqnarray}
where $q$ denotes the determinant of the metric tensor, $R$ is the Ricci scalar, $V$ is potential where $B_{\mu\nu}$ represents a self-interacting antisymmetric rank-2 tensor. The parameters $\xi_2$ and $\xi_3$ are the nonminimal coupling constants, while $\kappa = 8\pi G$ is the gravitational coupling constant. The term $\mathcal{L}(F)$ corresponds to the Lagrangian density of the  NLED  source, which is given by
\begin{equation}
{\cal L}(F)=\frac{3}{2sq^2}\left( \frac{\sqrt{2q^2F}}{1+\sqrt{2q^2F}}\right)^{5/2},
\label{nonl1}
\end{equation}
with \( s = \frac{q}{2M} \), where \( M \) and \( g \) are free parameters associated with the magnetic monopole's mass and charge, respectively. The energy-momentum tensor (EMT) is obtained by varying Eq.~(\ref{nonl1}) with respect to \( A_{\mu} \).

By performing a variation of Eq.~(\ref{action}) with respect to the metric \( g_{\mu\nu} \) and the electromagnetic potential \( A_{\mu} \), we obtain the modified Einstein field equations:
\begin{eqnarray}
&&G_{\mu\nu}\equiv R_{\mu\nu}-\frac{1}{2}g_{\mu\nu}R=T_{\mu\nu}^1+T_{\mu\nu}^2\\
&& \nabla_{a}\left(\frac{\partial {{\cal L}}}{\partial F}F^{a b}\right)=0\qquad \text{and} \qquad \nabla_{a}(* F^{ab})=0,
\label{eom}
\end{eqnarray}
where $T_{\mu\nu}^{(1)}$ and $T_{\mu\nu}^{(2)}$ denote the  EMTs corresponding to the Kalb-Ramond  field and the  NED source, respectively. Their explicit forms are given by
\begin{eqnarray}
T_{\mu\nu}^1&=& \frac{\xi_2}{\kappa}\Big[\frac{1}{2}g_{\mu\nu}B^{\alpha\gamma}B^{\beta}_{\gamma}] R_{\alpha\beta}-B^{\alpha}_{\mu} B{\beta} {\nu} R_{\alpha\beta}- B^{\alpha\beta}B_{\mu\beta}R_{\nu\beta}-B^{\alpha\beta}B_{\nu\beta}R_{\mu\alpha}\nonumber\\
&+& \frac{1}{2}D_{\alpha}D_{\mu}(B_{\nu\beta}B^{\alpha\beta})-\frac{1}{2}D^2(B{\alpha}_{\mu}B_{\alpha\nu})-\frac{1}{2}g_{\mu\nu}D_{\alpha}D_{\beta}(B^{\alpha\gamma}B^{\beta}_{\gamma})\Big],\label{t1}\\
T_{\mu\nu}^2&=&2\left[\frac{\partial {{\cal L}}}{\partial F}F_{\mu\sigma}F_{\nu}^{\sigma}-\tilde g_{\mu\nu}{{L(F)}}\right].
\label{eom1}
\end{eqnarray}
We consider a static, spherically symmetric solution to the vacuum Einstein field equations. Accordingly, the spacetime is described by the following line element:
\begin{equation}
ds^2 = -f(r)dt^2 +\frac{1}{f(r)}dr^2 + r^2d\Omega_2^2.
\end{equation}
The Kalb-Ramond  vacuum expectation value (VEV) ansatz is given by
\[
b_2 = \tilde{E}(r) \, dt \wedge \lambda \, dr,
\]
where the component \( b_{tr} = -\tilde{E}(r) \). The squared norm of the field is defined as
\[
b^2 = g^{\mu\nu} g^{\alpha\beta} b_{\mu\alpha} b_{\nu\beta}.
\]
Thus, the Kalb Ramond  VEV ansatz given in Eq.~(\ref{nonl1}) has a constant norm \( b^2 \) with respect to the metric specified in Eq.~(\ref{t1}), provided that 
\begin{equation}
\tilde E(r) = \sqrt{\frac {1}{2}}	|b|,
\end{equation}
where \( b \) is a constant. Observe that the function \( E(r) \) appearing in Eq.~\eqref{eom1} characterises a static, radial, pseudo-electric background field given by
\[
\tilde{E}^{\mu} = (0, \tilde{E}(r), 0, 0).
\]
The \((r, r)\)-component of the Einstein field equations is expressed as:
 \begin{equation}
 \frac{r^2\lambda}{2}f''(r)+(\lambda+1)rf(r) +f(r)-1=\frac{8M q^2}{(r^3+q^3)^{2}}, 
 \label{EERR}
 \end{equation}
where $\lambda \equiv b^2 \xi_2$. The equation of motion (\ref{EERR}) admits the following black hole solution:
\begin{equation}
ds^2=-\left(1-\frac{2M r^2}{r^3+q^3}+\frac{\gamma}{r^{2/\lambda}}\right)dt^2+\frac{dr^2}{\left(1-\frac{2M r^2}{r^3+q^3}+\frac{\gamma}{r^{2/\lambda}}\right)}+r^2d\Omega_2^2.
\label{bhs}`
\end{equation}
{The exact black hole solution described by Eq.~(\ref{bhs}) arises within the framework of general relativity coupled to a NLED source and includes Lorentz-violating contributions parameterized by $\gamma$ and $\lambda$. The spacetime geometry is fully characterized by four parameters: the mass $M$, the magnetic monopole charge $q$, and the Lorentz-violating parameters $\gamma$ and $\lambda$.

Several well-known black hole solutions can be recovered as particular limits of this geometry. In the absence of magnetic monopole charge ($q=0$), the solution reduces to the modified Kalb--Ramond black hole. In the limit $\gamma \to 0$ and $\lambda \to 0$ (equivalently $b^2 \to 0$ or $\xi_2 \to 0$), the solution approaches the Hayward black hole. Furthermore, when both $q \to 0$ and $\gamma \to 0$, the standard Schwarzschild black hole solution is recovered.

For $\gamma \geq 0$, $q=0$, and $\lambda = 1$ (with $0<\lambda\leq 2$), the obtained metric in Eq.~(\ref{bhs}) resembles the RN black hole solution \cite{25a,26a}. However, an important distinction arises in the electromagnetic sector. In this case, the pseudo-electric field becomes constant, $E(r)=|b|/\sqrt{2}$, which is consistent with an asymptotically flat spacetime containing a spacelike Lorentz-violating background field. Since a constant electric field cannot be generated by a localized charge distribution, the parameter $\gamma$ cannot be interpreted as an electric charge. Instead, it represents a Lorentz-violating hair associated with the black hole. Moreover, for $\lambda = 2$, the Lorentz-violating source no longer contains terms that depend on $\gamma$.

Finally, the energy conditions impose further constraints on the parameters. For $\lambda \leq 0$, both the weak and strong energy conditions are satisfied provided that $\gamma \leq 0$. In particular, for $\lambda = -1$, the solution satisfies the energy conditions only in the presence of a negative cosmological constant. The qualitative change in the behavior of the Lorentz-violating black hole solution when $\lambda$ varies from $1$ to $-1$ reflects the corresponding modification in the underlying source structure \cite{27a,28a}.

}

 \begin{figure*}[ht]
\begin{tabular}{c c c c}
\includegraphics[width=.52\linewidth]{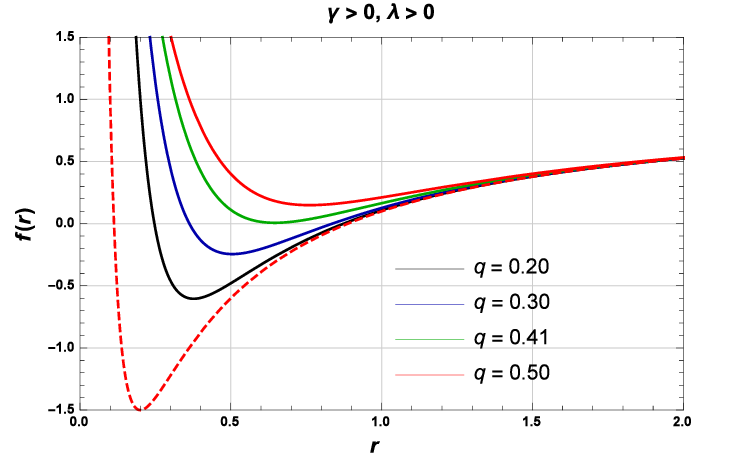}
\includegraphics[width=.52\linewidth]{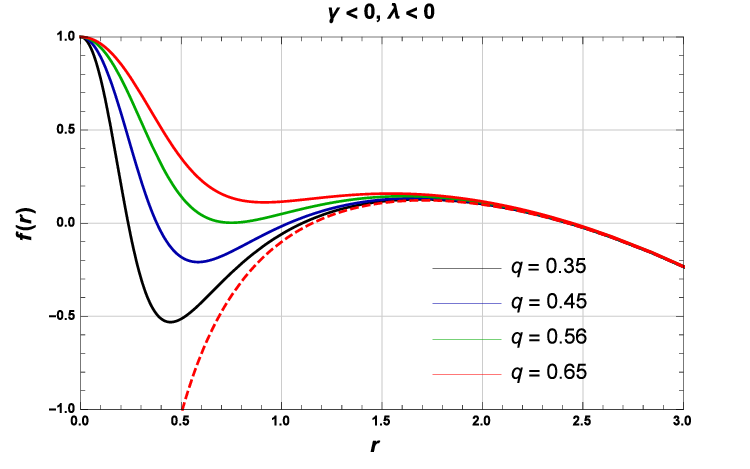}
\end{tabular}
\caption{Metric function $f(r)$ vs  $r$ for different values of magnetic monopole charge $q$ with fixed mass $M=0.50$. Left panel:  $\gamma = 0.1$   and $\lambda = 1$. Right panel:  $\gamma = -0.1$ and $\lambda=-1$.  }
\label{fig:1}
\end{figure*}

In Fig.~\ref{fig:1}, we present the metric function of Kalb–Ramond black hole solutions in an asymptotically flat spacetime for various values of charge parameters $q$. {The left and right panels display the behavior of the metric function for the parameter choices $\gamma = 0.1$, $\lambda = 1$, and $\gamma = -0.1$, $\lambda = -1$, respectively. The results indicate that, for $\gamma = 0.1$ and $\lambda = 1$, the black hole spacetime possesses two distinct horizons, namely an inner horizon and an outer (event) horizon. As the charge parameter $q$ decreases, the radius of the outer horizon gradually decreases. At the critical value $q = 0.41$, for fixed $\gamma = 0.1$ and $\lambda = 1$, the inner and outer horizons merge, leading to a degenerate horizon that corresponds to the extremal black hole configuration.}

{\bf In contrast, for $\gamma = -0.1$ and $\lambda = -1$, the solution exhibits three horizons. In this case, the Cauchy horizon and the event horizon coincide at the critical charge $q = 0.56$, resulting in an extremal configuration.} The black hole solutions cease to exist for values of $q$ exceeding this critical point.

\begin{figure*}[ht]
\begin{tabular}{c c c c}
\includegraphics[width=.55\linewidth]{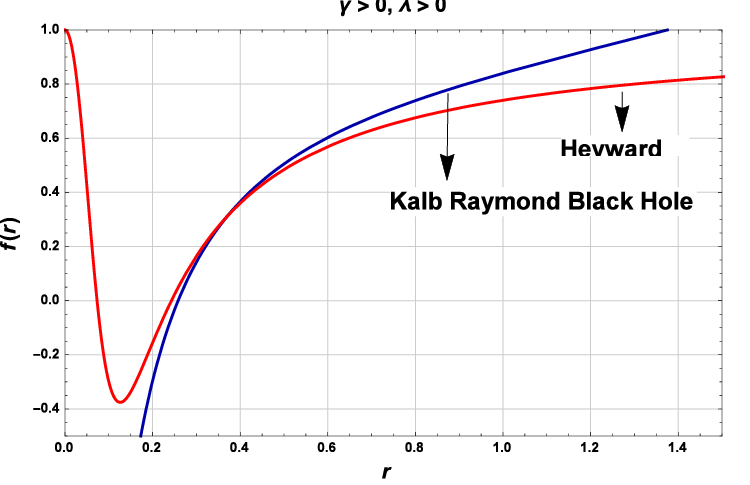}
\end{tabular}
\caption{Metric function $f(r)$ vs  $r$ for different values of magnetic monopole charge $q$ with fixed mass $M=0.13$.}
\label{fig:2}
\end{figure*}  

\section{Black Hole Thermodynamics}\label{sec3}
We determine the mass of the Kalb-Ramond  black hole spacetime in the presence of an NLED  source by solving equation~\eqref{bhs} under the condition $f(r_+) = 0$, which defines the event horizon, as follows: 
\begin{equation}\label{eq:12}
M=\frac{(q^{3}+r_{{+}}^{3})}{2 r_{{+}}^{2}} \Bigr[ 1+\frac{\gamma}{r_{{+}}^{\frac{2}{\lambda}}}+\frac{r_{{+}}^{2}}{l^{2}}\Bigr].     
\end{equation}
We compute the Hawking temperature of the black hole as
\begin{equation}\label{eq:14}
T_+= \frac{f'(r)}{4\pi}= \frac{1}{4 \pi}\left[ -\frac{4 M r_{{+}}}{q^{3}+r_{{+}}^{3}}+\frac{6 M r_{{+}}^{4}}{(q^{3}+r_{{+}}^{3})^{2}}-\frac{2 \gamma}{r_{{+}}^{\frac{2}{\lambda}} \lambda  r_{{+}}} \right].
\end{equation}
In the limit $l \to \infty$ and $q \to 0$, the equation yields the Hawking temperature corresponding to the asymptotically flat Kalb-Ramond  black hole spacetime in the absence of an NLED source
\begin{equation}\label{eq:15}
T_+= \frac{1}{4 \pi}\left[ -\frac{4 M }{r_{{+}}^{2}}+\frac{6 M }{r_{{+}}^{2}}-\frac{2 \gamma}{r_{{+}}^{\frac{2}{\lambda}} \lambda  r_{{+}}}\right].
\end{equation}

\begin{figure}
\begin{tabular}{c c}
 %         \begin{subfigure}{0.52\textwidth}
  %              \includegraphics[width=\linewidth]{t1.eps}
  %       \caption{$\gamma=0.1, \lambda=-1.$}
   %      \label{fig:7(a)}
   %  \end{subfigure}
     \begin{subfigure}{0.52\textwidth}
              \includegraphics[width=\linewidth]{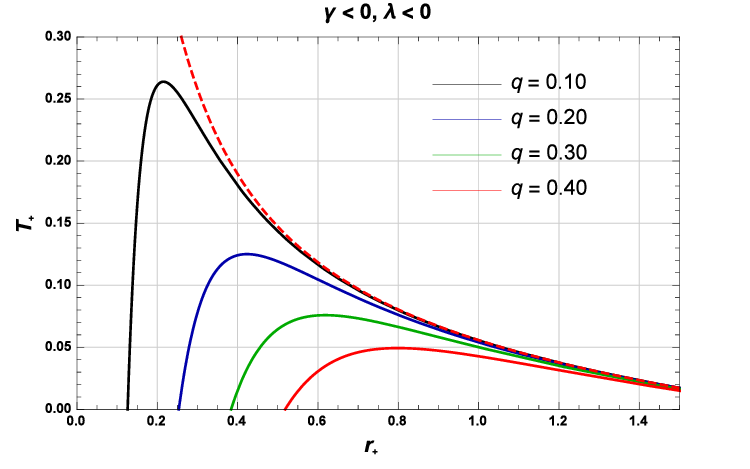}
         \caption{$\gamma=-0.1, \lambda=-1.$}
         \label{fig:7(b)}
     \end{subfigure}
         \begin{subfigure}{0.52\textwidth}
                 \includegraphics[width=\linewidth]{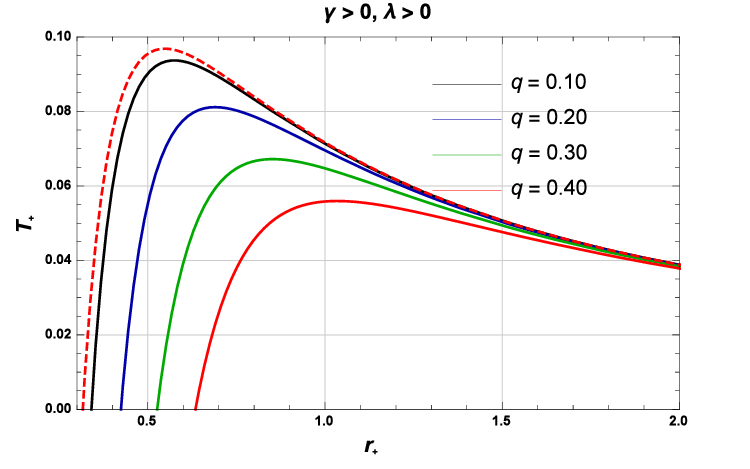}
         \caption{$\gamma= 0.1, \lambda= 1.$}
         \label{fig:7(c)}
     \end{subfigure}
   %  \begin{subfigure}{0.52\textwidth}
        %         \includegraphics[width=\linewidth]{t4.eps}
        % \caption{$\gamma=-0.1, \lambda= 1.$}
        % \label{fig:7(d)}
    % \end{subfigure}
     \end{tabular}
            \caption{The plot of temperature $T_+$ vs horizon radius  $r_+$ for different values of magnetic monopole charge  $q$.}
        \label{fig:7}
\end{figure}

In Fig.~\ref{fig:7}, we plot the Hawking temperature of a black hole in asymptotically flat spacetime. For positive values of $\gamma$ and $\lambda$ (Fig.~\ref{fig:7}), the Hawking temperature decreases monotonically with the horizon radius. When both Lorentz-violating parameters take negative values (Fig.~\ref{fig:7(b)}), the Hawking temperature first reaches a local maximum at $r_{+}^{a}$, followed by a local minimum at $r_{+}^{b}$, where $r_{+}^{b} > r_{+}^{a}$. As the horizon radius increases beyond $r_{+}^{b}$, the Hawking temperature rises slowly. When both Lorentz-violating parameters take positive values (Fig.~\ref{fig:7(b)}), the Hawking temperature decreases as the values of magnetic monopole charge increase.

%When one of the Lorentz-violating parameters assumes a negative value (Figs.~\ref{fig:7(a)} and \ref{fig:7(d)}), the Hawking temperature vanishes at a specific horizon radius, then gradually increases, and subsequently attains a local maximum at a critical horizon radius. Beyond this point, the Hawking temperature decreases with further increase in the horizon radius. 

Using the  first law of thermodynamics $C_+ dM = T_{+} \, dS$, \cite{ma} we obtain the entropy of the Kalb-Ramond AdS black hole coupled to  NLED as
\begin{equation}\label{eq:16}
    S=\pi r_+^{2}  -\frac{2 \pi  q^{3}}{r_+},
\end{equation}
where the integration constant is set to zero for simplicity. The influence of  NLED on the black hole entropy becomes evident in the preceding equation, as the entropy associated with NLED decreases by $2\pi q^3/r_+$. This entropy deviates from the standard area law. Since entropy must remain positive, the equation implies the following condition must be satisfied: $r_+ > 2^{1/3}q$.

 We now derive the first law of black hole thermodynamics in the extended phase 
space, along with the Smarr formula  \cite{Caldarelli:1999xj,Kastor:2009wy,Kubiznak:2016qmn}, wherein the pressure is treated as a thermodynamic variable.  The dimensional analysis yields: $[M] = L$, $[q] = L$, $[\gamma] = L^{2/\lambda}$. The thermodynamic quantities exhibit the following scaling behaviour:
$$
r_+ \to \alpha r_+, \quad q \to \alpha q,  \quad \gamma \to \alpha^{2/\lambda} \gamma, \quad \lambda \to \lambda, \quad M \to \alpha M, \quad S \to \alpha^2 S.
$$
\noindent Accordingly, the first law and the Smarr relation assume the following forms:
\begin{eqnarray}
    &&dM =T_+ dS +  \Phi_q dq + \Phi_{\lambda} d\lambda + \Gamma d\gamma,\\
  &&  M= 2T_+S+ \Phi_q q +\bigl( {2}/{\lambda}\bigl)  \Gamma \gamma.
    \label{eq:17}
\end{eqnarray}
Using the first law of black hole thermodynamics, we determine the potential and volume as
\begin{eqnarray}\label{eq:19}
    \Phi_q &=& \biggl( \frac{\partial{M}}{\partial{q}} \biggl)_{S,P,\lambda}= \frac{3q^{2}}{2 r_{{+}}^{2}}  \left(1+\frac{\gamma}{r_{{+}}^{\frac{2}{\lambda}}}+\frac{r_{{+}}^{2}}{l^{2}}\right),\qquad
    V  =  \biggl( \frac{\partial{M}}{\partial{P}} \biggl)_{S,q,\lambda}= \frac{4}{3} \pi(q^3+ r_{+}^3),\\
    \Phi_{\lambda}  &=&  \biggl( \frac{\partial{M}}{\partial{\alpha}} \biggl)_{S,P,q}= \frac{(q^{3}+r_{{+}}^{3}) r_{{+}}^{\frac{-2-2 \lambda}{\lambda}}}{2},\quad \text{and}\quad
\Gamma = \frac{\beta r_+^{\frac{-2-2 \lambda}{\lambda}} \ln  (r_+ ) (q^{3}+r_+^{3})}{\lambda^{2}}.
\end{eqnarray}
The black hole solution exhibits a well-defined specific heat
\begin{equation}
    C_+= T_+ \biggl( \frac{dS}{dT_+} \biggl)= \frac{8 \pi^{2} (q^{3}+r_{{+}}^{3})^{3} \lambda^{2} l^{2}}{A},
\end{equation}
where
\begin{eqnarray}
    A=&&\gamma   \left[10 q^{3} \Bigl(\lambda^{2}+\frac{3}{5} \lambda +\frac{4}{5}\Bigl)+r_+^3\Bigl(4-\lambda^{2}  \Bigl) \right]r_{{+}}^{\frac{-2+6 \lambda}{\lambda}}+\nonumber\\&&2 q^{6}  \gamma  (2+\lambda ) (1+\lambda ) r_{{+}}^{-\frac{2}{\lambda}}+2 \lambda^{2} \biggl( \Bigl( q^{6}+5 q^{3} r_{{+}}^{3}-\frac{1}{2} r_{{+}}^{6}\Bigl)\biggl).
\end{eqnarray}
\begin{figure}
\begin{tabular}{c c}
   %       \begin{subfigure}{0.5\textwidth}
   %             \includegraphics[width=\linewidth]{c1.eps}
     %   \caption{}
      %   \label{fig:8(a)}
   %  \end{subfigure}
     \begin{subfigure}{0.5\textwidth}
              \includegraphics[width=\linewidth]{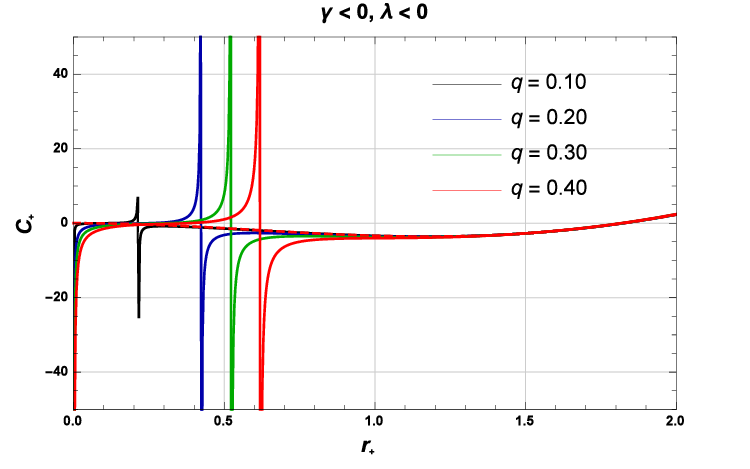}
         \caption{$\gamma=-0.10$ and $\lambda=-1.0$}
         \label{fig:8(b)}
     \end{subfigure}
         \begin{subfigure}{0.5\textwidth}
                 \includegraphics[width=\linewidth]{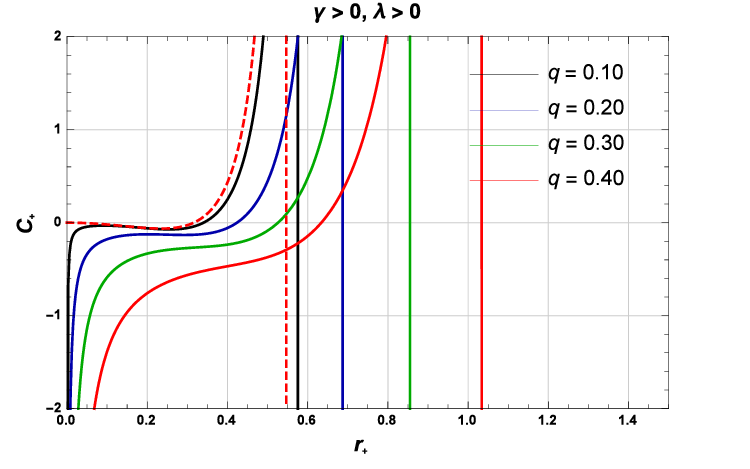}
         \caption{$\gamma=0.10$ and $\lambda=1.0$}
         \label{fig:8(c)}
     \end{subfigure}
    % \begin{subfigure}{0.5\textwidth}
            %     \includegraphics[width=\linewidth]{c4.eps}
       %  \caption{}
        % \label{fig:8(d)}
     %\end{subfigure}
     \end{tabular}
            \caption{The plot of heat capacity ($C_+$) vs horizon radius  ($r_+$ )for different values of the ($\gamma$) and ($\lambda$) parameter with a fixed value of black hole mass ($M=1$) and magnetic monopole charge ($q$).}
        \label{fig:8}
\end{figure}

{In Fig.~\ref{fig:8}, the behaviour of the heat capacity of the Kalb-Ramond  black hole is illustrated.  Fig.~\ref{fig:8(b)} illustrates the heat capacity behavior when one Lorentz-violating parameter is negative. In this case, the heat capacity is discontinuous at the Hawking temperature. For values of $r_+ > 1.8$, the heat capacity remains continuous and positive, indicating thermodynamic stability for the corresponding black hole solutions. Fig.~\ref{fig:8(c)} depicts the thermodynamic behaviour for positive values of both the  Lorentz-violating parameters. The heat capacity diverges once at the minimum of the Hawking temperature (Fig.~\ref{fig:8(c)}). The divergence in heat capacity increases with the magnetic monopole charge.

}

To check the global stability of the black hole, we study the Gibbs free energy, defined as 
\begin{equation}
    G_+= M-T_+S.
\end{equation}
Exploiting equations \eqref{eq:12}, \eqref{eq:14} and \eqref{eq:15}, the above relation leads to
\begin{eqnarray}
G_+= &&-\frac{ \pi( r_+^{3}  -2  q^{3})}{(q^{3} +r_+^{3})r_+^2\lambda}\left( \gamma  (\lambda -2) r_+^{\frac{-2+3 \lambda}{\lambda}}-2 q^{3}  \gamma  (1+\lambda ) r_+^{-\frac{2}{\lambda}}-4 \lambda (2q^{3}-{r_+^{3}})  \right) +\frac{q^{3}+r_+^{3}}{2 r_+^{2}}\left(1+\gamma r_+^{-\frac{2}{\lambda}} \right).
\end{eqnarray}
\begin{figure}
\begin{tabular}{c c}
       %   \begin{subfigure}{0.5\textwidth}
             %   \includegraphics[width=\linewidth]{11a.eps}
       %  \caption{}
        % \label{fig:11(a)}
    % \end{subfigure}
     \begin{subfigure}{0.5\textwidth}
              \includegraphics[width=\linewidth]{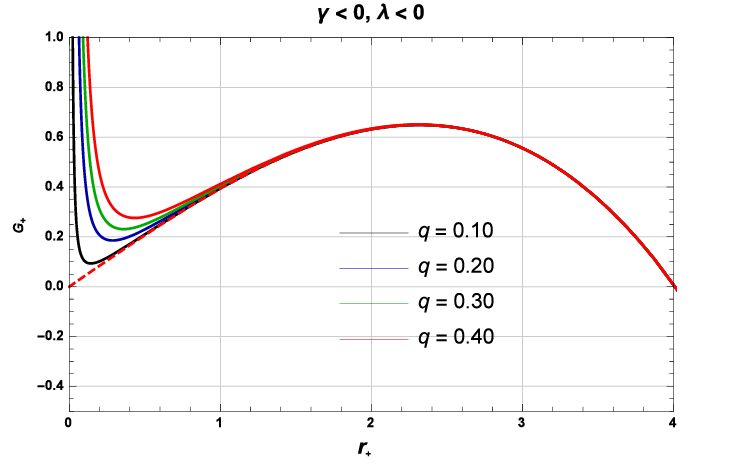}
         \caption{$\gamma=-0.10$ and $\lambda=-1.0$}
         \label{fig:11(b)}
     \end{subfigure}
         \begin{subfigure}{0.5\textwidth}
                 \includegraphics[width=\linewidth]{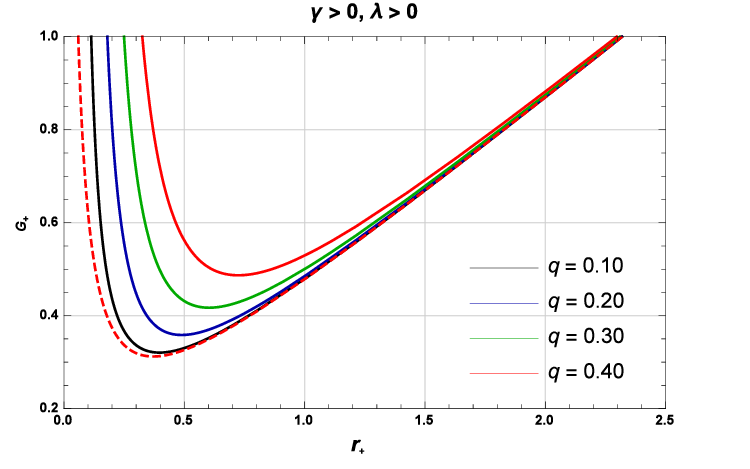}
         \caption{$\gamma=0.10$ and $\lambda=1.0$}
         \label{fig:11(c)}
     \end{subfigure}
     %\begin{subfigure}{0.5\textwidth}
             %   \includegraphics[width=\linewidth]{11d.eps}
        % \caption{}
         %\label{fig:11(d)}
    % \end{subfigure}
     \end{tabular}
\caption{The plot of Gibbs free energy ($G_+$) vs  $r_+$ for different values of the Kalb-Ramond  parameter with a fixed value of black hole mass ($M=1$) and magnetic monopole charge ($q$).}
\label{fig:11}
\end{figure}

When $G_+ \leq 0$, the black hole configuration exhibits global thermodynamic stability. In Fig.~\ref{fig:11}, the behaviour of the Gibbs free energy corresponding to asymptotically flat black hole solutions is illustrated. Fig.~\ref{fig:11} illustrates the variation of the Gibbs free energy for different Lorentz-violating scenarios.

{
In Figs.~\ref{fig:11(b)}, the Gibbs free energy profile is presented for negative values of the 
Lorentz-violating parameters. The Gibbs free energy initially decreases with increasing horizon radius 
$r_+$, attaining a local minimum. At this minimum, the associated Hawking temperature exhibits a local 
maximum, as is evident in Figs.~\ref{fig:7(b)} and \ref{fig:7(c)}. Beyond this point, the Gibbs free 
energy reverses its trend and increases with $r_+$ and reaches the maximum values, then begins to decrease.

In  Fig.~\ref{fig:11(c)}, the Gibbs free energy initially decreases with increasing horizon radius $r_+$, attaining a local minimum. Beyond this point, the Gibbs free energy reverses its trend and increases with $r_+$.  In Fig.~\ref{fig:11(b)}, a local minimum and maximum are observed; after that, the free energy transitions to a monotonically decreasing regime.}
 
%%%%%%%%%%%%%%%%%%%%%%%%%%%%%%%%%%%%%%%%%%%%%%%%%%%%%%%%%%%%%%%%%%%%%
\section{$P-v$  Criticality and Phase Transitions}\label{sec4}
In this section, we investigate the thermodynamic phase structure of Hayward black holes in the presence of the Kalb-Ramond  field. Utilizing the expression for the Hawking temperature given in Eq.~\eqref{eq:14}, we derive the corresponding black hole equation of state
\begin{equation}\label{eq:30}
P= \frac{T_+}{v} -\frac{1}{2 \pi  v^{2}} +\frac{8 T_+ q^{3}}{v^{4}} +\frac{8 q^{3}}{\pi  v^{5}}+\frac{8 q^{3} \gamma}{\pi  v^{5}  (v/2)^\frac{2}{\lambda}}+\frac{8 q^{3} \gamma}{\pi  \lambda  v^{5} (v/2)^\frac{2}{\lambda}}-\frac{\gamma}{2 \pi  v^{2} (v/2)^\frac{2}{\lambda}}+\frac{\gamma}{\pi  \lambda  v^{2} (v/2)^\frac{2}{\lambda}},
\end{equation}
where $v=2r_+$. The critical point is characterised by the pressure \( P \) satisfying the following conditions at the point of inflection \cite{4,5}
\begin{equation}\label{eq:31}
    \Biggl( \frac{\partial{P}}{\partial{v}}\Biggl)_{T_{c},v_{c}}= \Biggl( \frac{\partial^2{P}}{\partial{v}^2}\Biggl)_{T_{c},v_{c}}=0.
\end{equation}
Using equation \eqref{eq:30} and  \eqref{eq:31} we obtain equation for critical volume 
\begin{eqnarray}
&&-8 \gamma (\lambda +1) (\lambda -2) \biggr[ -64 q^{3} (-1+\lambda ) v_{c}^{\frac{-2+3 \lambda}{\lambda}}+v_{c}^{\frac{-2+6 \lambda}{\lambda}} (2+\lambda )\biggr]    2^{\frac{2-3 \lambda}{\lambda}}\nonumber\\
&&
+16 \gamma (5\lambda +2) (2\lambda +1) (\lambda +1)       q^{3}  [  v_{c}^{3} +8      q^{3}     ] 4^{\frac{1}{\lambda}} v_{c}^{-\frac{2}{\lambda}}  +1280 q^{6} \lambda^{3}+224 q^{3} \lambda^{3} v_{c}^{3}-\lambda^{3} v_{c}^{6} = 0.\label{eq:32}
\end{eqnarray}
We numerically solved Eq.~\eqref{eq:32} to determine the critical points. For the case $q=0$, the critical ratio $\rho_c = \frac{3}{8}$ corresponds precisely to the behaviour of the Van der Waals fluid, indicating an exact analogy with black hole thermodynamics in this regime. As the parameter $q$ increases from zero, the critical radius and the critical ratio $\rho_c$ exhibit an increasing trend. In contrast, the critical pressure and critical temperature decrease monotonically with increasing $q$. These results reveal the significant influence of $q$ on the thermodynamic phase structure of the black hole system.

The thermodynamic equation of state associated with the black hole configuration is expressed as
\begin{equation}\label{eq:33}
P=-\frac{\lambda  (4 \pi  T_+ q^{3} r_{+} +4 \pi  T_+ r^{4}+2 q^{3}-r_{+}^{3}) r_{+}^{\frac{2}{\lambda}}}{8 \pi  (2 q^{3} \lambda -r_{+}^{3} \lambda +2 q^{3}+2 r_{+}^{3})},
\end{equation}
where we used $\gamma=-8 \pi P$. At critical points, the following conditions must be satisfied:
\begin{equation}\label{eq:34}
\Biggl( \frac{\partial{P}}{\partial{v}}\Biggl)_{T_{c},v_{c}}= \Biggl( \frac{\partial^2{P}}{\partial{v}^2}\Biggl)_{T_{c},v_{c}}=0.
\end{equation}
{Utilising Eqs.~\eqref{eq:33} and \eqref{eq:34}, we determine the critical thermodynamic parameters corresponding to the coupling constants $\lambda = -1.0$
\begin{eqnarray}
    r_c= (14+6 \sqrt{6})^{\frac{1}{3}} q,\qquad T_c=\frac{3+2 \sqrt{6}}{4 \pi q (3+\sqrt{6}) (14+6 \sqrt{6})^{{1}/{3}} }, \quad \text{and} \quad P_c=\frac{3(5+2 \sqrt{6}) }{16 \pi q^2 2^{{2}/{3}} (7+3 \sqrt{6})^{\frac{5}{3}}   (3+\sqrt{6})}
\end{eqnarray}

and $\lambda = -0.50$ as:
\begin{eqnarray}
   r_c=2.6397q, \qquad T_c=\frac{ 0.0294}{q }, \quad\text{and}\quad P_C= \frac{0.00002}{q^4}.
\end{eqnarray}

}

%\begin{center}
	%\begin{table}[h]
		%\begin{center}
			%\begin{tabular}{l| l| r lr l r l r}
			%	\hline
				%\multicolumn{1}{c|}{ }&\multicolumn{1}{c|}{ $\lambda=-1.0$  }&\multicolumn{1}{c}{$\lambda=-0.5$}\\
			%	\hline
				
% $r_c$ & $(14+6 \sqrt{6})^{\frac{1}{3}} q $ & $2.6397 q$ \\ 

% $T_c$ & $\frac{3+2 \sqrt{6}}{4 \pi q (3+\sqrt{6}) (14+6 \sqrt{6})^{{1}/{3}} }$ & $\frac{ 0.0294}{q }$ \\ 

% $P_c$ & $\frac{3(5+2 \sqrt{6}) }{16 \pi q^2 2^{{2}/{3}} (7+3 \sqrt{6})^{\frac{5}{3}}   (3+\sqrt{6})}$ & $\frac{0.00002}{q^4}  $ \\ 

			%	\hline
			%\end{tabular}
		%\end{center}
		% \caption{Critical parameters.}
        % \label{tab}
%	\end{table}
%\end{center}

\begin{figure}
\begin{tabular}{c c}
          \begin{subfigure}{0.5\textwidth}
                \includegraphics[width=\linewidth]{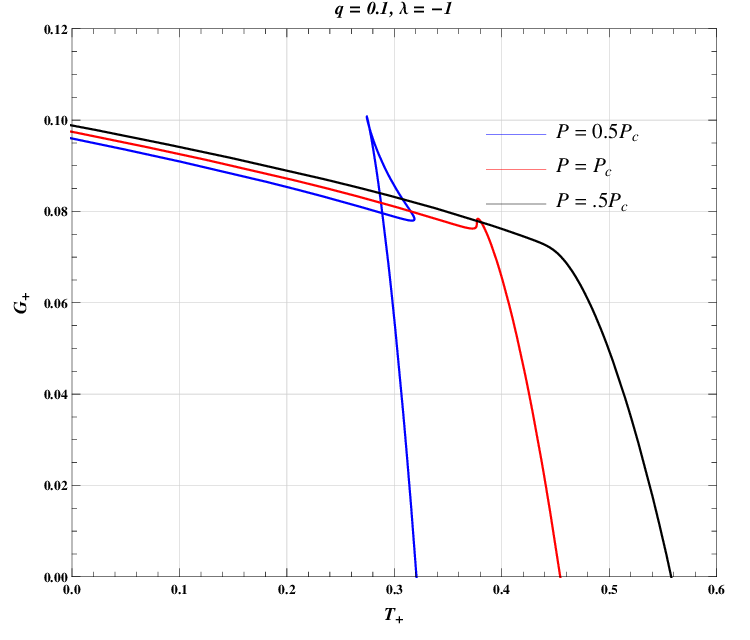}
         \caption{}
         \label{fig:99(a)}
     \end{subfigure}
     \begin{subfigure}{0.5\textwidth}
              \includegraphics[width=\linewidth]{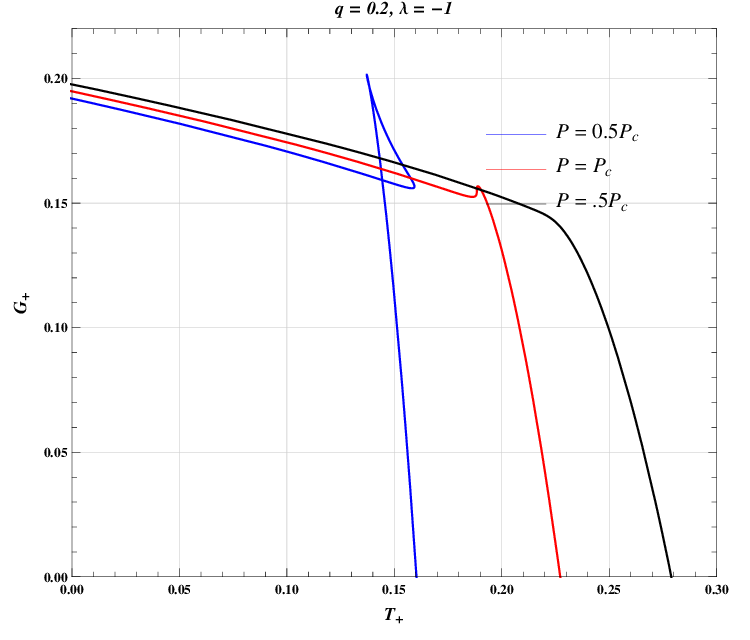}
         \caption{}
         \label{fig:99(b)}
     \end{subfigure} 
     \end{tabular} 
\caption{The plot $G_+$ vs  $T_+$ for different values  magnetic monopole charge ($q$) with fixed values of $\lambda =-1$.}
\label{fig:14}
\end{figure}
 
In Fig.~\ref{fig:14}, we present the $G_{+}$–$T_{+}$ diagram for two distinct values of the parameter $\lambda$, considering three thermodynamic regimes: subcritical pressure ($P<P_c$), critical pressure ($P=P_c$), and supercritical pressure ($P>P_c$). For $P<P_c$, the $G_{+}$–$T_{+}$ diagram exhibits a characteristic swallowtail structure, signifying the presence of a first-order phase transition between small black hole and large black hole phases. This thermodynamic behaviour is analogous to Van der Waals fluids' liquid-gas transition. At the critical pressure $P=P_c$, the swallowtail structure vanishes, indicating the termination of the first-order phase transition and the emergence of a second-order critical point. For $P>P_c$, the diagram reveals a smooth, monotonic profile, and no phase transition is observed, thereby confirming the single-phase nature of the system in the supercritical regime.

\section{Conclusion} \label{sec5}

This study examined the black hole solutions in $AdS$ spacetime, incorporating a Kalb-Ramond  field coupled to  NLED. We have derived the governing action, leading to the modified Einstein field equations, and have obtained an exact black hole solution characterised by mass \(M\), magnetic monopole charge \(q\), and Lorentz-violating parameters \(\gamma\) and \(\lambda\).
Our analysis has revealed the presence of two distinct horizons: an inner and an outer horizon. We have observed that as the monopole charge \(q\) decreases, the radius of the outer horizon increases. The two horizons merge at a critical value of \(q = 0.56\), resulting in a single degenerate horizon. The black hole solutions cease for values of \(q\) exceeding this critical point.
Furthermore, we have demonstrated that in the limiting case where the Lorentz-violating parameters vanish, the solution reduces to the modified Kalb-Ramond  and Hayward solutions. Additionally, by setting specific parameter values, the solution has been shown to reduce to the RN–$AdS$ black hole and the Schwarzschild–$AdS$ black hole.

This work's exact black hole solution has provided new insights into the effects of a Kalb-Ramond  field and NLED in $AdS$ spacetime. Our findings have implications for the study of black holes in modified gravity theories and may contribute to a deeper understanding of the interplay between field theories and gravitational solutions.
In the thermodynamic analysis, we computed the Hawking temperature and examined the behaviour of entropy, specific heat, and Gibbs free energy for various parameter configurations. Our findings have revealed that the presence of the NLED source introduces significant modifications to the thermodynamic properties of the Kalb-Ramond  black hole. Specifically, we have observed that the Hawking temperature exhibits non-monotonic behaviour, with the possibility of local maxima and minima, depending on the values of the Lorentz-violating parameters. Additionally, we have found that the entropy deviates from the standard area law, and the specific heat can become negative, indicating thermodynamic instability in certain regimes.
Moreover, we have derived the first law of black hole thermodynamics in the extended phase space, including the Smarr relation, treating pressure as a thermodynamic variable. We have confirmed that the standard first law and the Smarr formula remain valid for the charged Kalb-Ramond  black holes, even in the presence of the NLED source. Our analysis has also shown that the Gibbs free energy exhibits non-monotonic behavior, with the possibility of local minima and maxima, signaling the presence of first-order phase transitions.

In conclusion, our study has provided a comprehensive thermodynamic analysis of the Kalb-Ramond  black hole in the presence of an NLED source. We have demonstrated that including the NLED source results in significant modifications to the black hole's thermodynamic properties, offering new insights into the effects of Lorentz violation on black hole thermodynamics.

\begin{acknowledgements}  
This research was funded by the Science Committee of the Ministry of Science and Higher Education of the Republic of Kazakhstan (Grant No. AP23487178).
  DVS would like to thank the Council of Science and Technology (CST), Uttar Pradesh, for the project (grant no. CST/D-828).
\end{acknowledgements}
\section*{ Data availability statement}
There is no data in this manuscript. 
%%%%%%%%%%%%%%%%%%%%%%%%%%%%%%%%%%%%%%%%%%%%%%%%%%%%%%%%%%%%%%%%%%%%


\begin{thebibliography}{99}
\bibitem{01} V.A. Kostelecky and S. Samuel, ``Spontaneous breaking of Lorentz symmetry in string theory", {Phys.Rev. D} {39} (1989) 683.
\bibitem{02} J. Alfaro, H.A. Morales-Tecotl and L.F. Urrutia, ``Loop quantum gravity and light propagation", {Phys.Rev. D} {65} (2002) 103509.
\bibitem{03} P. Horava, ``Quantum Gravity at a Lifshitz Point", {Phys. Rev.} {79} (2009) 084008.
\bibitem{04} S.M. Carroll, J.A. Harvey, V.A. Kostelecky, C.D. Lane and T. Okamoto, ``Noncommutative field theory and Lorentz violation", {Phys. Rev. Lett.} {87} (2001) 141601. 
\bibitem{05} T. Jacobson and D. Mattingly, ``Gravity with a dynamical preferred frame", {Phys. Rev. D} {64} (2001) 024028.
\bibitem{06} A.G. Cohen and S.L. Glashow, ``Very special relativity", {Phys. Rev. Lett.} {97} (2006) 021601.
\bibitem{07}B. Altschul, Q. G. Bailey, and V. A. Kostelecky, ``Lorentz violation with an antisymmetric tensor", Phys. Rev. D 81 (2010),
065028.
\bibitem{7} M. Green, J. Schwarz, E. Witten, (1985) ``Superstring Theory", Volume 2 (Cambridge University Press,
Cambridge).
\bibitem{8} M. Kalb and P. Ramond, ``Classical direct interstring action", Phys. Rev. D 9, 2273 (1974).
\bibitem{9} F. Hehl, P. von der Heyde, G. Kerlick and J. Nester, ``General relativity with spin and torsion: Foundations and prospects", Rev. Mod. Phys. 48, 393 (1976).
\bibitem{10} F. Hehl, P. von der Heyde, G. Kerlick and J. Nester, ``Metric affine gauge theory of gravity: Field equations, Noether identities, world spinors, and breaking of dilation invariance", Phys. Rept. 258, 1 (1995).
  

\bibitem{Born:1934ji}
M. Born. ``On the quantum theory of the electromagnetic field”, Proc. Roy. Soc. Lond. A 143.849 (1934), pp. 410–437.
\bibitem{Born:1934gh}
 M. Born and L. Infeld. ``Foundations of the new field theory”,  Proc. Roy. Soc. Lond. A 144.852 (1934), pp. 425–451.
 \bibitem{Born:1935ap}
 M. Born and L. Infeld. ``On the quantization of the new field equations. II”, Proc. Roy. Soc. Lond. A 150.869 (1935), pp. 141–166.
\bibitem{Infeld:1936wzo}
L. Infeld. ``The new action function and the unitary field theory”, Proc. Cambridge Phil. Soc. 32.1 (1936), pp. 127–137.
\bibitem{Infeld:1936frv}
L. Infeld. ``A new group of action functions in the unitary field theory. II”, Proc. Cambridge Phil. Soc. 33.1 (1937), pp. 70–78.
\bibitem{Ayon-Beato:2000mjt}
Eloy Ayon-Beato and Alberto Garcia. ``The Bardeen model as a nonlinear magnetic monopole”, Phys. Lett. B 493 (2000), pp. 149–152.
\bibitem{Ayon-Beato:1999qin}
 Eloy Ayon-Beato and Alberto Garcia. ``Nonsingular charged black hole solution for nonlinear source”, Gen. Rel. Grav. 31 (1999), pp. 629–633.
\bibitem{Ayon-Beato:1998hmi}
 Eloy Ayon-Beato and Alberto Garcia. ``Regular black hole in general relativity coupled to nonlinear electrodynamics”, Phys. Rev. Lett. 80 (1998), pp. 5056–5059.
\bibitem{bardeen1968proceedings}
James M Bardeen, Proceedings of the International Conference GR5. 1968.

%%%%%%%%%%%%%%%%%%%%%%%%%%%%
\bibitem{Singh:2025svv}
D.~V.~Singh, S.~Upadhyay, Y.~Myrzakulov, K.~Myrzakulov, B.~Singh and M.~Kumar,
``Thermodynamic behavior and phase transitions of black holes with a cloud of strings and perfect fluid dark matter,'' Nucl. Phys. B {1016} (2025), 116915.
\bibitem{Kumar:2024zqf}
A.~Kumar, D.~V.~Singh and S.~Upadhyay,
``Impact of Perfect Fluid Dark Matter on the Thermodynamics of AdS Ay\'on-Beato-Garc\'Ia Black Holes,''
JHAP {4} (2024) no.4, 85-99.
\bibitem{Kumar:2024cnh}
A.~Kumar, D.~V.~Singh and S.~Upadhyay,
``Ay\'on\textendash{}Beato\textendash{}Garc\'\i{}a black hole coupled with a cloud of strings: Thermodynamics, shadows and quasinormal modes,''
Int. J. Mod. Phys. A {39} (2024) no.31, 2450136.
\bibitem{Sudhanshu:2024wqb}
H.~K.~Sudhanshu, D.~V.~Singh, S.~Upadhyay, Y.~Myrzakulov and K.~Myrzakulov,
``Thermodynamics of a newly constructed black hole coupled with nonlinear electrodynamics and cloud of strings,''
Phys. Dark Univ. {46} (2024), 101648.
\bibitem{Singh:2024jgo}
B.~Singh, D.~Veer Singh and B.~Kumar Singh,
``Thermodynamics, phase structure and quasinormal modes for AdS Heyward massive black hole,''
Phys. Scripta {99} (2024) no.2, 025305.
\bibitem{Kumar:2023cmo}
A.~Kumar, D.~V.~Singh, Y.~Myrzakulov, G.~Yergaliyeva and S.~Upadhyay,
``Exact solution of Bardeen black hole in Einstein\textendash{}Gauss\textendash{}Bonnet gravity,''
Eur. Phys. J. Plus {138} (2023) no.12, 1071.
\bibitem{Singh:2023czv}
B.~Singh, B.~K.~Singh and D.~V.~Singh,
``Thermodynamics, phase structure of Bardeen massive black hole in Gauss-Bonnet gravity,''
Int. J. Geom. Meth. Mod. Phys. {20} (2023) no.08, 2350125.
\bibitem{Hayward:2005gi}
S.~A.~Hayward,
``Formation and evaporation of regular black holes,''
Phys. Rev. Lett. {96} (2006), 031103.
\bibitem{Singh:2022xgi}
D.~V.~Singh, S.~G.~Ghosh and S.~D.~Maharaj,
``Exact nonsingular black holes and thermodynamics,''
Nucl. Phys. B {981} (2022), 115854.
\bibitem{Kumar:2020bqf}
A.~Kumar, D.~V.~Singh and S.~G.~Ghosh,
``Hayward black holes in Einstein\textendash{}Gauss\textendash{}Bonnet gravity,''
Annals Phys. {419} (2020), 168214.
\bibitem{Singh:2019wpu}
D.~V.~Singh, S.~G.~Ghosh and S.~D.~Maharaj,
``Bardeen-like regular black holes in $5D$ Einstein-Gauss-Bonnet gravity,''
Annals Phys. {412} (2020), 168025.
\bibitem{Kumar:2018vsm}
A.~Kumar, D.~V. Singh and S.~G.~Ghosh,
`$D$-dimensional Bardeen-AdS black holes in Einstein-Gauss-Bonnet theory,''
Eur. Phys. J. C {79} (2019) no.3, 275.
\bibitem{Ghosh:2018bxg}
S.~G.~Ghosh, D.~V.~Singh and S.~D.~Maharaj,
``Regular black holes in Einstein-Gauss-Bonnet gravity,''
Phys. Rev. D {97} (2018) no.10, 104050.
 \bibitem{2a}
 S.H. Hendi, G.Q. Li, J.X. Mo, S. Panahiyan, and B. Eslam Panah,
 New perspective for black hole thermodynamics in Gauss-Bonnet Born-Infeld massive gravity, 
Eur. Phys. J. C 76 (2016) 571.
 \bibitem{6a}
S.H. Hendi, B. Eslam Panah and S. Panahiyan, 
``Black Hole Solutions in Gauss- Bonnet- Massive Gravity in the Presence of Power- Maxwell Field", 
Fortschr. Phys. (Prog. Phys.) 2018 (2018) 1800005.
\bibitem{Vishvakarma:2023tnl}
B.~K.~Vishvakarma, D.~V.~Singh and S.~Siwach,
``Parameter estimation of the Bardeen-Kerr black hole in cloud of strings using shadow analysis,''
Phys. Scripta  {99} (2024) no. 2, 025022.
\bibitem{sud00}P. Paul, S. Upadhyay, Y. Myrzakulov, D. V. Singh and K. Myrzakulov, ``More Exact Thermodynamics of Nonlinear Charged AdS Black Holes in 4D Critical Gravity", Nucl. Phys. B 993 (2023) 116259.
\bibitem{sud01}Y. Myrzakulov, K. Myrzakulov, S. Upadhyay and D. V. Singh, 
``Quasinormal Modes and Phase Structure of Regular  Einstein-Gauss-Bonnet Black Holes",
International Journal of Geometric Methods in Modern Physics 20, 2350121 (2023).
 \bibitem{ma}
 M. Ma and R. Zhao, Class. Quantum Grav. 31 245014 (2014).
 \bibitem{dvs20}
  D. V. Singh and S. Siwach,``Thermodynamics and P-v criticality of Bardeen-AdS Black Hole in $4D$ Einstein-Gauss-Bonnet Gravity" Physics Letter B {\bf 808} (2020) 135658.
\bibitem{1}
D.~V.~Singh, V.~K.~Bhardwaj and S.~Upadhyay,
``Thermodynamic properties, thermal image and phase transition of Einstein-Gauss-Bonnet black hole coupled with nonlinear electrodynamics,''
Eur. Phys. J. Plus {137} (2022) no.8, 969.
\bibitem{sud}A. Kumar, D. V. Singh and S. Upadhyay, ``Ayón–Beato–García black hole coupled with a cloud of strings: Thermodynamics, shadows and quasinormal modes", Int. J. Mod. Phys. A 39 (2024) 2450136.
\bibitem{sud1}H. K. Sudhanshu, D. V. Singh, S. Upadhyay, Y. Myrzakulov and K. Myrzakulov, 
``Thermodynamics of a newly constructed black hole coupled with nonlinear electrodynamics and cloud of strings",
Physics of the Dark Universe 46 (2024) 101648.
\bibitem{sud2}D. V. Singh, A. Shukla and S. Upadhyay, ``Quasinormal modes, shadow and thermodynamics of black holes coupled with nonlinear electrodynamics and cloud of strings",
Annals of Physics 447 (2022) 169157.
%%%%%%%%%%%%%%%%%%%%%%%%%
\bibitem{Bekenstein:1973ur}
J. D. Bekenstein. ``Black holes and entropy”, Phys. Rev. D 7 (1973), pp. 2333–2346.
\bibitem{Bardeen:1973gs}
J. M. Bardeen, B. Carter, and S. W. Hawking. ``The Four laws of black hole mechanics”, Commun. Math. Phys. 31 (1973), pp. 161–170.
\bibitem{Hawking:1982dh}
 S. W. Hawking and Don N. Page. ``Thermodynamics of Black Holes in anti-de Sitter Space”, Commun. Math. Phys. 87 (1983), p. 577.
 \bibitem{25a}
H. Reissner, ``Über die Eigengravitation des elektrischen Feldes nach der Einsteinschen Theorie"  Ann. Physik 50, 106–120 (1916).
\bibitem{26a}
 G. Nordstrom, ``On the Energy of the Gravitation field in Einstein's Theory" Proc. Kon. Ned. Akad. 20, 1238–1245 (1918).
\bibitem{27a}
L. A. Lessa, J. E. G Silva, R. V. Maluf, C. A. S. Almeida, ``Modified black hole solution with a background Kalb–Ramond field, Eur. Phys. J. C (2020) 80:335.
\bibitem{28a}
K. K. Nandi, R. N. Izmailov, R. Kh. Karimov, A. A. Potapov, ``On the Kalb–Ramond modified Lorentz violating hairy black holes and Thorne’s hoop conjecture" Eur. Phys. J. C (2023) 83:984.

\bibitem{Caldarelli:1999xj}
 M. M. Caldarelli, G. Cognola, and D. Klemm, ``Thermodynamics of Kerr-Newman-AdS black holes and conformal field theories," Class. Quant. Grav. 17 (2000).


%%%%%%%%%%%%%%%%%%%%%%%%%%%%%%%%%

\bibitem{Kastor:2009wy}
D.~Kastor, S.~Ray and J.~Traschen,
``Enthalpy and the Mechanics of AdS Black Holes", 
Class. Quant. Grav. {26} (2009), 195011.
\bibitem{Kubiznak:2016qmn}
D.~Kubiznak, R.~B.~Mann and M.~Teo,
 ``Black hole chemistry: thermodynamics with Lambda", 
Class. Quant. Grav. {34} (2017) no.6, 063001.
\bibitem{4}
D. Kastor, S. Ray, and J. Traschen, ``Enthalpy and the Mechanics of AdS Black Holes”, Class. Quant. Grav. 26 (2009), p. 195011.
\bibitem{5}
D. Kubiznak, R. B. Mann, and M. Teo, ``Black hole chemistry: thermodynamics with Lambda", Class. Quant. Grav. 34.6 (2017), p. 063001.
\end{thebibliography}
\end{document}